\documentclass[aps,prd,reprint,groupedaddress,nofootinbib,longbibliography]{revtex4-2}

\usepackage[utf8]{inputenc}
\usepackage{amsmath,amssymb,mathtools}
\usepackage{graphicx}
\usepackage{bm}
\usepackage{mathrsfs}
\usepackage{tensor}
\usepackage{hyperref}
\usepackage{xcolor}

\newcommand{\dd}{\mathrm d}
\newcommand{\TT}{\mathrm{TT}}
\newcommand{\GR}{\mathrm{GR}}
\newcommand{\phib}{\bar\phi}
\newcommand{\Wphi}{\mathcal W_\phi}
\newcommand{\PTT}{\mathcal P_{\TT}^{[1,1]}}

\newcommand{\calT}{\mathcal T}
\newcommand{\calE}{\mathcal E}
\newcommand{\calF}{\mathcal F}
\newcommand{\calA}{\mathcal A}

\begin{document}

\title{Gravitational Waves as Thermodynamic Shear Excitations in Scalar--Tensor Gravity}

\author{David S. Pereira}
\email{djpereira@ciencias.ulisboa.pt}
\affiliation{Departamento de F\'isica, Faculdade de Ci\^encias da Universidade de Lisboa,
Campo Grande, Edif\'icio C8, P-1749-016 Lisboa, Portugal}
\affiliation{Instituto de Astrof\'isica e Ci\^encias do Espa\c{c}o,
Faculdade de Ci\^encias da Universidade de Lisboa,
Campo Grande, Edif\'icio C8, P-1749-016 Lisboa, Portugal}

\date{\today}

\begin{abstract}
Can a propagating spin--2 mode participate in a local thermodynamic process without assigning an autonomous entropy or a unique local energy to the wave itself? We show that this occurs as a local thermodynamic work process in Jordan-frame scalar--tensor gravity, where a timelike scalar gradient defines an internal gravitational thermodynamic medium. Building on the first-order TT deformation--response relation, we identify the local shear power $\mathcal W_\phi=\pi_{ab}^{(\phi)}\sigma^{ab}$ from scalar-frame traction mechanics and show that its leading pure-TT contribution is second order, gauge invariant, and enters the exact scalar-sector energy balance. We further show that the same effective anisotropic stress controls the scalar--tensor modification of the tensor-friction term, so the local shear-work process is directly encoded in gravitational-wave propagation. In geometric optics the corresponding ratio of shear work to GR-normalized tensor energy is controlled by the running effective Planck mass, providing a local thermodynamic interpretation of the standard modified gravitational-wave luminosity-distance law. Finally, we extend this shear-excitation mechanism to nonminimally coupled scalars, metric $f(R)$ gravity, and viable luminal Horndeski theories, and identify the condition required for the same interpretation in more general tensor sectors.
\end{abstract}

\maketitle

\section{Introduction}
\label{sec:intro}

The relation between gravitation and thermodynamics has repeatedly revealed structures that are not apparent when the gravitational field equations are viewed only as equations of motion. Black-hole mechanics, horizon entropy, local Clausius relations, and nonequilibrium extensions of Einstein gravity all suggest that gravitational dynamics can, in suitable settings, be organized in terms of thermodynamic variables and balance laws \cite{Bekenstein:1973ur,Hawking:1975vcx,Wald:1993nt,Jacobson:1995ab,Eling:2006aw,Chirco:2010se}. Gravitational radiation occupies a more complex position. Its energy can be characterized asymptotically and in controlled short-wavelength limits \cite{Bondi:1962px,Sachs:1962wk,Ashtekar:1981bq,Isaacson:1968zza,Burnett:1989gp}, but these constructions do not by themselves endow a propagating spin--2 perturbation with a local thermodynamic work law.

In this work we ask a related question: Can gravitational waves participate in a thermodynamic process when additional gravitational degrees of freedom provide the required state variables, fluxes, constitutive stresses, and balance laws? More concretely, can a propagating spin--2 mode mechanically deform an internal gravitational medium and exchange energy with it through a well-defined local stress power? If such an exchange exists, is it also encoded in the propagation of the gravitational wave? This formulation deliberately does not assume an autonomous gravitational-wave entropy or a unique covariant local wave energy.

Scalar--tensor gravity provides a minimal realization of this possibility. When the scalar gradient is timelike, the Jordan-frame scalar contribution admits an exact imperfect-fluid decomposition whose constitutive relations, heat flux, anisotropic stress, and balance laws define an effective first-order thermodynamic medium \cite{Eckart:1940te,Faraoni:2018qdr,Faraoni:2021lfc,Faraoni:2021jri,Faraoni:2023hwu,Giardino:2023ygc}. The framework has been explored in cosmology, exact scalar--tensor spacetimes, induced gravity, and viable Horndeski theories \cite{Giusti:2022tgq,Giardino:2023qlu, Giusti:2021sku,Miranda:2024dhw,Gallerani:2024gdy, Giusti:2026ymb,Faraoni:2025dex,Faraoni:2025alq}. At the same time, recent analyses show that a simple Eckart interpretation is not automatic in more general nonminimal or multiscalar theories \cite{Miranda:2022wkz,Pereira:2025dmk,Pereira:2026hpi}. This makes scalar--tensor gravity useful not only as a thermodynamic model, but also as a setting in which one can separate a robust local mechanical statement from stronger assumptions about entropy or thermal closure.

The direct precursor of the present work is Ref.~\cite{Pereira:2026xog}. There it was shown that a pure transverse--traceless (TT) perturbation of a homogeneous shear-free background leaves the scalar-frame expansion, acceleration, heat flux, and the perturbation of the invariant product $\mathcal K\mathcal T$ unchanged at first order, while generating shear. The first-order TT anisotropic stress is proportional to that shear. The wave therefore excites a conjugate deformation--response pair in the internal scalar medium without activating the heat or volume-expansion channels at the same order. This identifies a possible thermodynamic role for the tensor mode, but not yet a work process: mechanical power is bilinear in stress and deformation rate and must be selected by a local mechanical principle. The work~\cite{Pereira:2026xog} therefore provides the kinematical and constitutive starting point for the present analysis, but does not by itself establish a local work law. The present paper addresses this missing step by determining the mechanically relevant power directly from the scalar-frame material deformation and stress.

Furthermore, there is also an independent propagation-side clue. Effective anisotropic stress and nonstandard gravitational-wave propagation are linked in broad classes of modified gravity \cite{Saltas:2014dha}. In scalar--tensor and Horndeski theories, the same nonminimal coupling that enters the trace-free scalar response also changes the tensor friction term through the running effective Planck mass \cite{Bellini:2014fua,Kobayashi:2011nu,Deffayet:2011gz,Lagos:2019kds}. Integrated along the ray, this modification changes the gravitational-wave luminosity distance \cite{Belgacem:2017ihm,Belgacem:2018,Amendola:2017ovw}. The multimessenger constraint from GW170817/GRB170817A further selects luminal subclasses of Horndeski and related theories \cite{Baker:2017hug,Creminelli:2017sry,Ezquiaga:2017ekz,Sakstein:2017xjx,Langlois:2017dyl}. These results suggest that the local constitutive response of the scalar medium and the propagation anomaly of the tensor mode may be two manifestations of the same coupling.

The purpose of this paper is to answer both questions within a single framework: whether a propagating TT mode activates a well-defined local thermodynamic shear-work channel of the internal scalar medium, and whether the same constitutive coupling is independently encoded in the modified propagation of the gravitational wave. We first identify precisely how a TT perturbation deforms the scalar thermodynamic medium. We then localize the scalar-sector traction power and show that its unique trace-free contribution is $\Wphi\equiv\pi_{ab}^{(\phi)}\sigma^{ab}$, where $\pi_{ab}^{(\phi)}$ is the effective scalar anisotropic stress and $\sigma_{ab}$ is the shear tensor of the scalar-gradient congruence.  For a homogeneous shear-free background, the leading pure-TT work is a second-order, gauge-invariant bilinear and enters the exact scalar-sector energy balance. We next return to tensor propagation and show that the same effective
anisotropic stress entering this local shear work also controls the
tensor-friction modification, establishing a direct source--power
correspondence. The geometric-optics relation, the standard-siren modification, and the extensions to Brans--Dicke, nonminimally coupled scalars, metric $f(R)$ gravity, viable luminal Horndeski, and more general tensor sectors are then developed as consequences and tests of this thermodynamic shear-excitation mechanism.

Throughout the paper we work in the Jordan matter frame, which is operationally distinguished when ordinary matter is minimally coupled. The effective-fluid decomposition is not claimed to be conformal-frame invariant. Coordinate gauge invariance and conformal-frame dependence will be kept conceptually separate. We also make no claim that the gravitational wave itself carries an autonomous thermodynamic entropy or that the GR-normalized tensor amplitude defines a unique local wave energy. The thermodynamic statement established below concerns the local constitutive shear-work channel of the internal scalar-gradient medium activated by the TT deformation. The expressions thermodynamic work and thermodynamic exchange refer to mechanical work occurring within this already defined effective scalar thermodynamic medium. They do not imply an autonomous entropy for the gravitational wave or identify the shear-work channel with ordinary positive viscous entropy production; stronger entropy statements require additional thermodynamic assumptions that will be introduced separately.

The paper is organized as follows. In Sec.~\ref{sec:scalarTT} we introduce the scalar thermodynamic medium and recall how a TT perturbation excites its shear and anisotropic-stress channels while leaving the first-order heat and expansion channels unchanged. In Sec.~\ref{sec:exchange} we pass from this deformation--response pair to a genuine thermodynamic exchange by deriving the local scalar-frame traction power, identifying the shear-work scalar, and establishing its second-order perturbative and energetic properties. Section~\ref{sec:prop} then shows that the same constitutive stress is encoded in tensor propagation and develops the geometric-optics and standard-siren consequences of the resulting source-power correspondence. In Sec.~\ref{sec:families} we test how far this mechanism extends beyond Bergmann--Wagoner gravity, considering nonminimally coupled scalars, metric $f(R)$ gravity, viable luminal Horndeski theories, and a generic tensor sector. The main implications, limitations, and the broader question of whether an analogous thermodynamic channel can arise in general relativity are discussed in Sec.~\ref{sec:discussion}. 

\section{Scalar-tensor thermodynamics and its TT deformation}
\label{sec:scalarTT}

To answer the first part of the question, we must identify what the tensor mode actually changes in the internal scalar medium. We will begin in Bergmann--Wagoner gravity \cite{Bergmann:1968ve,Wagoner:1970vr}, where both the effective thermodynamic variables and the tensor dynamics can be written in closed form. In a vacuum propagation region the action is given by
\begin{equation}
S=\frac{1}{16\pi}\int\!\dd^4x\sqrt{-g}
\left[
\phi R-\frac{\omega(\phi)}{\phi}(\nabla\phi)^2-2V(\phi)
\right],
\label{eq:action}
\end{equation}
with signature $(-+++)$. Variation with respect to the scalar gives
\begin{equation}
\left[2\omega(\phi)+3\right]\Box\phi=-\frac{\dd\omega}{\dd\phi}\nabla_\mu\phi\nabla^\mu\phi
+2\phi\frac{\dd V}{\dd\phi}
-4V(\phi),
\label{eq:scalar_field}
\end{equation}
whereas the metric equation can be written in Einstein-like form as
\begin{equation}
G_{ab}=8\pi T_{ab}^{(\phi)},
\label{eq:einsteinlike}
\end{equation}
with
\begin{align}
8\pi T_{\mu\nu}^{(\phi)}={}&\frac{\omega(\phi)}{\phi^2}\left(\nabla_\mu\phi\nabla_\nu\phi-\frac12g_{\mu\nu}\nabla_\alpha\phi\nabla^\alpha\phi\right)\nonumber\\
&+\frac1\phi\left(\nabla_\mu\nabla_\nu\phi-g_{\mu\nu}\Box\phi\right)-\frac{V(\phi)}{\phi}g_{\mu\nu}.
\label{eq:Tphi}
\end{align}
It is this Einstein-like representation that makes the scalar-gradient fluid description possible. We restrict throughout to the positive-effective-Planck-mass sector
\begin{equation}
\phi>0,
\label{eq:positive_phi}
\end{equation}
so that the tensor kinetic coefficient is positive and $M_*^2=\phi/(8\pi)>0$. In particular, the background satisfies $\phib(t)>0$ throughout the propagation region considered below.

We will perturb a spatially flat Friedmann--Lema\^itre--Robertson--Walker (FLRW) background according to
\begin{equation}
\phi=\phib(t),\qquad
g_{ij}=a^2(t)\left(\delta_{ij}+\varepsilon\chi_{ij}^{\TT}\right),
\label{eq:metric}
\end{equation}
where
\begin{equation}
\partial^i\chi_{ij}^{\TT}=0,
\quad
\chi^i{}_{i}=0.
\end{equation}
Spatial indices on $\chi_{ij}^{\TT}$ are raised and lowered with $\delta_{ij}$.
For perturbative bookkeeping one has
\begin{equation}
Q=\bar Q+\varepsilon\delta Q+\frac{\varepsilon^2}{2}\delta^{(2)}Q+\cdots,
\label{eq:expansion}
\end{equation}
and we will denote by $[1,1]$ the terms formed from two first-order TT fields and by $[2]$ the terms linear in genuine second-order perturbations \cite{Bruni:1996im}. At first order one can represent the pure tensor sector with $g_{00}=-1$ and $g_{0i}=0$ \cite{Bardeen:1980kt}. At second order, however, tensor--tensor couplings source scalar, vector, and tensor metric perturbations; in a general gauge this means that lapse, shift, scalar, and spatial-metric corrections cannot consistently be assumed to vanish \cite{Noh:Hwang:2004,Zhang:Qin:Wang:2017}.

When the background equations are satisfied, the quadratic tensor action is given by
\begin{equation}
S_T^{(2)}=\frac{1}{64\pi}\int\!\dd^4x\,a^3\phib\left[\dot\chi_{ij}^{\TT}\dot\chi_{\TT}^{ij}-a^{-2}(\partial_k\chi_{ij}^{\TT})^2\right],
\label{eq:tensor_action}
\end{equation}
and one has the propagation equation
\begin{equation}
\ddot\chi_{ij}^{\TT}+\left(3H+\frac{\dot\phib}{\phib}\right)\dot\chi_{ij}^{\TT}-a^{-2}\partial^2\chi_{ij}^{\TT}=0.
\label{eq:tensor_eq}
\end{equation}
Moreover, the tensor kinetic coefficient is proportional to the effective Planck mass, $M_*^2=\phib/(8\pi)$. Whenever $H\neq0$, it is convenient to introduce
\begin{equation}
\alpha_M\equiv\frac{\dd\ln M_*^2}{\dd\ln a}=\frac{\dot\phib}{H\phib},
\label{eq:alphaM}
\end{equation}
so that the tensor friction term can be written as $(3+\alpha_M)H$.

We will now recall the scalar-frame variables that are needed below. If $\nabla_a\phi$ is timelike and future directed, one defines
\begin{equation}
u_a\equiv\frac{\nabla_a\phi}{X},
\qquad
X\equiv\sqrt{-\nabla_a\phi\nabla^a\phi},
\qquad
h_{ab}=g_{ab}+u_au_b.
\label{eq:u}
\end{equation}
On the homogeneous background one has $u^a=(1,0,0,0)$ and $X=-\dot\phib$. Together with $\phib>0$, we therefore restrict to the future-directed decreasing-$\phi$ branch, $\dot\phib<0$, for which the effective thermal coefficient introduced below is positive \cite{Giusti:2022tgq}.

Relative to $u^a$, the exact decomposition of $T_{ab}^{(\phi)}$ is
\begin{equation}
T_{ab}^{(\phi)}=\rho_\phi u_au_b+p_\phi h_{ab}+2u_{(a}q_{b)}^{(\phi)}+\pi_{ab}^{(\phi)}.
\label{eq:fluid}
\end{equation}
The kinematical decomposition of the scalar-frame congruence is given by
\begin{equation}
\nabla_a u_b=-u_a a_b+\frac13\Theta h_{ab}+\sigma_{ab},
\end{equation}
where
\begin{equation}
a_b\equiv u^c\nabla_cu_b,
\qquad
\Theta\equiv\nabla_a u^a.
\end{equation}

The scalar $\Theta$ measures the local volume-expansion rate of the scalar-frame congruence, while $\sigma_{ab}$ is its symmetric, spatial, trace-free shear tensor. Direct projection of Eq.~\eqref{eq:Tphi} gives the exact constitutive identities \cite{Faraoni:2021lfc,Faraoni:2021jri,Faraoni:2023hwu,Giardino:2023ygc}
\begin{align}
q_a^{(\phi)}&=D_a(\mathcal K\mathcal T)=-\mathcal K\mathcal T\,a_a,
\nonumber\\
\pi_{ab}^{(\phi)}&=\mathcal K\mathcal T\,\sigma_{ab},
\qquad
\mathcal K\mathcal T=\frac{X}{8\pi\phi}>0.
\label{eq:constitutive}
\end{align}
These are the relations underlying the first-order thermodynamic description of the scalar sector \cite{Faraoni:2021lfc,Faraoni:2021jri,Faraoni:2023hwu,Giardino:2023ygc}. Notice that the gravitational equations determine only the product $\mathcal K\mathcal T$; a separate conductivity and temperature require additional thermodynamic input. The scalar sector is already thermodynamic in the Eckart sense: the field equations define an effective medium with constitutive fluxes and stresses. The TT mode acquires a thermodynamic role by exciting its shear channel.

We will consider now the TT perturbation of this medium. The complete first-order calculation was carried out in Ref.~\cite{Pereira:2026xog}. For a pure TT mode on the homogeneous shear-free background one has
\begin{equation}
\delta X\big|_{\TT} = \delta(\mathcal K\mathcal T)\big|_{\TT} = \delta\Theta\big|_{\TT} = \delta a_i\big|_{\TT} = \delta q_i^{(\phi)}\big|_{\TT} =0,
\label{eq:TTzero}
\end{equation}
while the shear and anisotropic stress are
\begin{align}
\delta\sigma_{ij}\big|_{\TT} &= \frac{a^2}{2}\dot\chi_{ij}^{\TT},
\nonumber\\
\delta\sigma^i{}_{j}\big|_{\TT} &= \frac12\dot\chi^i{}_{j},
\nonumber\\
\delta\pi^i{}_{j}{}^{(\phi)}\big|_{\TT} &= \frac{\overline{\mathcal K\mathcal T}}{2}\dot\chi^i{}_{j}.
\label{eq:TTshearstress}
\end{align} 

The tensor mode does not perturb the scalar state variable or the heat channel at first order. Instead, it excites a purely trace-free deformation and the corresponding trace-free stress. To make the mechanical analogy more transparent, consider
\begin{equation}
\epsilon^i{}_{j}\equiv\frac12\chi^i{}_{j}
\end{equation}
be the tensor strain. Then
\begin{equation}
\delta\sigma^i{}_{j}\big|_{\TT}=\dot\epsilon^i{}_{j},
\qquad
\delta\pi^i{}_{j}{}^{(\phi)}\big|_{\TT}=\overline{\mathcal K\mathcal T}\,\dot\epsilon^i{}_{j}.
\label{eq:strainlaw}
\end{equation}
One therefore has, at first order, a genuine deformation--response pair. This is the kinematic content of calling the wave a {shear excitation}: the tensor mode changes the shape of the scalar-frame material element and induces the conjugate trace-free stress, while the first-order heat and expansion channels remain unexcited. What is still missing is the thermodynamic exchange itself. We will now determine whether this deformation--response pair carries a mechanically selected local power.

\section{From TT deformation to thermodynamic exchange}
\label{sec:exchange}

We can now address the first part of the central question. A deformation--response pair becomes a thermodynamic work exchange when the associated constitutive stress performs a well-defined local mechanical power in the already identified effective thermodynamic medium. We will therefore consider a small scalar-frame material element and localize the traction power acting on its deformation. This step determines the relevant scalar from mechanics itself, independently of the scalar--tensor constitutive law that later evaluates its magnitude, and it remains valid for the genuinely accelerated scalar-frame congruence.

\subsection{Localization of the stress power}

Consider now a small spatial region $\mathcal V_\tau$ made of neighboring worldlines of the scalar-frame congruence $u^a$. The spatial projection of the effective stress tensor is
\begin{equation}
S_{ab}^{(\phi)} \equiv h_a{}^ch_b{}^dT_{cd}^{(\phi)} = p_\phi h_{ab}+\pi_{ab}^{(\phi)}.
\label{eq:spatialstress}
\end{equation}
For an outward unit normal $n^a$, the traction is
\begin{equation}
t^a=S_{(\phi)}^{ab}n_b.
\end{equation}

Let $\xi^a$ be a connecting vector between neighboring scalar-frame worldlines, Lie transported by the congruence,
\begin{equation}
\mathcal L_u\xi^a=0.
\label{eq:lie_dragged_connector}
\end{equation}
For an accelerated congruence, a Lie-dragged connector that is initially spatial need not remain orthogonal to $u^a$. We therefore introduce its instantaneous spatial projection
\begin{equation}
\ell^a\equiv h^a{}_b\xi^b,
\qquad
u_a\ell^a=0,
\label{eq:spatial_connector}
\end{equation}
and define the relative spatial velocity of neighboring worldlines by
\begin{equation}
v^a
\equiv
h^a{}_b u^c\nabla_c\ell^b.
\label{eq:relativev_def}
\end{equation}
Defining $s\equiv u_b\xi^b$, so that $\xi^a=\ell^a-su^a$, the Lie-transport condition gives $u^c\nabla_c\xi^a=\xi^c\nabla_cu^a$, and therefore
\begin{align}
h^a{}_b u^c\nabla_c\xi^b
&=
B_b{}^a\ell^b-s a^a,
\nonumber\\
h^a{}_b u^c\nabla_c(su^b)
&=
s a^a,
\label{eq:connector_cancellation}
\end{align}
where
\begin{equation}
B_{ab}
\equiv
h_a{}^ch_b{}^d\nabla_cu_d.
\label{eq:Bdef}
\end{equation}
The two acceleration terms cancel in the derivative of
$\ell^a=\xi^a+s u^a$, so that
\begin{equation}
v^a
=
B_b{}^a\ell^b.
\label{eq:relativev_transposed}
\end{equation}
Because $u_a\propto\nabla_a\phi$, the scalar-frame congruence is hypersurface orthogonal.
Its vorticity therefore vanishes and
\begin{equation}
B_{ab}=B_{ba}=\frac13\Theta h_{ab}+\sigma_{ab}.
\label{eq:Bsymmetry}
\end{equation}
Consequently,
\begin{equation}
v^a=B_b{}^a\ell^b=B^a{}_b\ell^b.
\label{eq:relativev}
\end{equation}

Thus the standard local deformation map remains valid even when the scalar-frame congruence is accelerated, and the apparent index transposition is removed by the vorticity-free character of the scalar-gradient flow. In the local affine approximation around the central worldline, one therefore has
\begin{equation}
D_{(a}v_{b)}
=
B_{(ab)}
=
B_{ab}
=
\frac13\Theta h_{ab}+\sigma_{ab}
=
\frac12\mathcal L_u h_{ab}
\equiv
\mathcal D_{ab}.
\label{eq:Dab}
\end{equation}
This is the relativistic rate-of-deformation tensor of the scalar-frame material element; the thread-based continuum viewpoint is particularly natural for such a construction
\cite{Aoki:2022ipw}.

The construction above is purely kinematic and does not require $a_a=0$.
This is essential in the present application because the scalar-gradient congruence is generically accelerated whenever the effective heat channel is nontrivial.

The signed boundary power is given by
\begin{equation}
\mathcal P_{\partial\mathcal V}=\int_{\partial\mathcal V_\tau}\dd A\,(S_{(\phi)}^{ab}n_b)v_a,
\label{eq:boundarypower}
\end{equation}
and the spatial divergence theorem gives
\begin{equation}
\mathcal P_{\partial\mathcal V}=\int_{\mathcal V_\tau}\dd V\left[(D_bS_{(\phi)}^{ab})v_a+S_{(\phi)}^{ab}D_bv_a\right].
\label{eq:divpower}
\end{equation}
The first term is the power of the resultant force on the bulk motion.
Because $S_{ab}^{(\phi)}$ is symmetric, the internal deformation power density is
\begin{equation}
\mathcal P_{\rm mech}^{(\phi)}
=
S_{(\phi)}^{ab}D_{(a}v_{b)}
=
S_{(\phi)}^{ab}\mathcal D_{ab},
\label{eq:Pmech1}
\end{equation}
and using Eq.~\eqref{eq:spatialstress} and the orthogonal trace decomposition one gets
\begin{equation}
\mathcal P_{\rm mech}^{(\phi)}=p_\phi\Theta+\pi_{ab}^{(\phi)}\sigma^{ab}.
\label{eq:Pmech2}
\end{equation}

We have therefore arrived at the relativistic counterpart of the standard continuum-mechanical split of stress power into volumetric and deviatoric pieces.
In particular, Eq.~\eqref{eq:Pmech2} is the mechanical origin of the work scalar
\begin{equation}
\Wphi
\equiv
\pi_{ab}^{(\phi)}\sigma^{ab}
=
\frac12\pi_{(\phi)}^{ab}\mathcal L_u h_{ab}.
\label{eq:W}
\end{equation}
It is the unique trace-free part of the localized traction power.
It is therefore selected before the scalar--tensor constitutive relation is used.
Only after the mechanical scalar has been identified do we substitute Eq.~\eqref{eq:constitutive}, obtaining
\begin{equation}
\Wphi=\mathcal K\mathcal T\,
\sigma_{ab}\sigma^{ab}=\frac{\pi_{ab}^{(\phi)}\pi_{(\phi)}^{ab}}
{\mathcal K\mathcal T}.
\label{eq:Wforms}
\end{equation}
Equation~\eqref{eq:Wforms} is  the standard deviatoric stress-power scalar once a stress tensor and a rate-of-deformation tensor have been specified. The point established here is not a new algebraic identity of relativistic fluid mechanics. Rather, the traction derivation fixes which stress of the effective gravitational sector is mechanically conjugate to the scalar-frame deformation, before the constitutive relation is invoked. This distinction is important in the present context because an effective source obtained by rearranging a field equation need not, by itself, be the traction stress of an independently defined medium.

The sign convention is inherited from the boundary traction. Positive $S_{(\phi)}^{ab}\mathcal D_{ab}$ denotes work exerted by the effective scalar stress on the deformation. Thus $p_\phi>0$ and $\Theta>0$ correspond to positive outward pressure work during expansion, and the trace-free term uses the same convention. Accordingly, positive $\Wphi$ denotes work performed by the scalar constitutive stress on the shear deformation; it is not defined as positive work deposited into the scalar sector. As will follow directly from the scalar energy balance below, the same contribution therefore enters the scalar-sector internal-energy change with the opposite sign, $-\Wphi$. The TT perturbation supplies the shear deformation that activates this work channel. With the local mechanical scalar fixed, we can now return to the tensor perturbation.

\subsection{Pure-TT work at leading nonvanishing order}

Because the FLRW background is shear free,
$\bar\sigma_{ab}=0$ and $\bar\Wphi=0$.
The first nonvanishing pure-TT contribution is bilinear in the first-order stress and deformation rate:
\begin{align}
\PTT
&\equiv
(\Wphi)_{\TT}^{[1,1]}
\nonumber\\
&=
\delta\pi^i{}_{j}{}^{(\phi)}\big|_{\TT}\,
\delta\sigma^j{}_{i}\big|_{\TT}
\nonumber\\
&=
\overline{\mathcal K\mathcal T}\,
\delta\sigma^i{}_{j}\big|_{\TT}
\delta\sigma^j{}_{i}\big|_{\TT}
\nonumber\\
&=
\frac{\overline{\mathcal K\mathcal T}}{4}
\dot\chi_{ij}^{\TT}\dot\chi_{\TT}^{ij}.
\label{eq:Pmain}
\end{align}
On the future-directed decreasing-$\phi$ branch,
\begin{equation}
\overline{\mathcal K\mathcal T}
=
-\frac{\dot\phib}{8\pi\phib}>0,
\label{eq:KTbar}
\end{equation}
so
\begin{equation}
\PTT
=
-\frac{\dot\phib}{32\pi\phib}
\dot\chi_{ij}^{\TT}\dot\chi_{\TT}^{ij}
\ge0.
\label{eq:Pphi}
\end{equation}
The positivity is branch dependent; fundamentally $\Wphi$ is a signed stress-power density. Equation~\eqref{eq:Pmain} is the precise sense in which the tensor mode becomes a thermodynamic shear excitation: the wave does not acquire an autonomous thermodynamic state, but its TT strain rate activates the deviatoric work channel of the internal scalar medium.

For a small material cell over which the wave and background vary negligibly, the instantaneous TT shear power is
\begin{equation}
\dot W_{\rm shear}^{\rm cell}
\simeq
V_{\rm cell}\,\PTT.
\label{eq:cellpower}
\end{equation}
This makes explicit that $\PTT$ has the operational meaning of local work per proper volume per scalar-frame proper time.

It is useful to make the physical size of this effect more explicit.
We will consider now a locally monochromatic wave and write
\begin{equation}
\chi_{ij}^{\TT}
=
\sum_{\lambda=+,\times}
e_{ij}^{\lambda}
A_\lambda\cos(\theta+\varphi_\lambda),
\qquad
e_{ij}^{\lambda}e^{ij}_{\lambda'}
=
2\delta_{\lambda\lambda'},
\label{eq:polarization}
\end{equation}
where the amplitudes $A_\lambda$ and relative phases $\varphi_\lambda$ vary slowly, and the local physical frequency is
$\omega=-\mathscr D_u\theta$.
To leading geometric-optics order,
\begin{equation}
\left\langle
\dot\chi_{ij}^{\TT}\dot\chi_{\TT}^{ij}
\right\rangle_{\rm cyc}
=
\omega^2\sum_\lambda A_\lambda^2,
\label{eq:dotchiavg}
\end{equation}
and hence
\begin{equation}
\left\langle\PTT\right\rangle_{\rm cyc}
=
\frac{\overline{\mathcal K\mathcal T}}{4}
\omega^2
\sum_\lambda A_\lambda^2.
\label{eq:Pavg}
\end{equation}
There is no polarization splitting in this parity-even scalar--tensor sector.
For $\overline{\mathcal K\mathcal T}>0$ the cycle-averaged stress power has a definite sign because the constitutive stress is proportional to the deformation rate rather than the deformation itself.

\subsection{Cyclic work and the rheological character of the TT channel}
\label{subsec:cyclicwork}

The local power law contains information that is not present in the first-order deformation--response relation alone. To make this explicit, we can consider one wave cycle over which the background quantities and amplitudes can be treated as constant. With $\epsilon^i{}_j=\chi^i{}_j/2$, define the shear-work one-form per proper volume by
\begin{equation}
\delta w_{\rm sh}
\equiv
\pi^{(\phi)i}{}_{j}\,\dd\epsilon^j{}_{i}.
\label{eq:workoneform}
\end{equation}
For any closed strain history $\gamma$ with $\epsilon^i{}_j(\tau_f)=\epsilon^i{}_j(\tau_i)$, the constitutive law gives the sign-definite cycle functional
\begin{equation}
\Delta w_{\rm sh}[\gamma]
\equiv
\oint_\gamma \pi^{(\phi)i}{}_{j}\,\dd\epsilon^j{}_{i}
=
\int_{\tau_i}^{\tau_f}\!\dd\tau\,
\mathcal K\mathcal T\,
\dot\epsilon^i{}_j\dot\epsilon^j{}_i
\ge 0,
\label{eq:cyclicworkgeneral}
\end{equation}
on the branch $\mathcal K\mathcal T>0$, with equality only for a shear-free history.  Thus a closed deformation cycle can perform finite net work even though the strain returns to its initial value.  No conservative strain state function $U_{\rm sh}(\epsilon)$ can reproduce this response, because such a function would give $\oint\dd U_{\rm sh}=0$.

For the monochromatic wave in Eq.~\eqref{eq:polarization}, Eq.~\eqref{eq:cyclicworkgeneral} reduces at leading geometric-optics order to
\begin{equation}
\Delta w_{\rm sh}^{\rm cyc}
=
\frac{\pi}{2}\,
\overline{\mathcal K\mathcal T}\,
\omega\sum_{\lambda}A_\lambda^2.
\label{eq:cyclicwork}
\end{equation}
The TT channel is therefore not merely a relabeling of an elastic deformation: it is a rate-dependent, path-dependent work channel of the scalar medium.

The same statement is transparent in frequency space. With the convention $\epsilon^i{}_j\propto e^{-i\omega\tau}$, Eq.~\eqref{eq:strainlaw} gives
\begin{equation}
\widetilde{\pi}^{(\phi)i}{}_{j}
=
-i\omega\,
\overline{\mathcal K\mathcal T}\,
\widetilde{\epsilon}^{i}{}_{j}.
\label{eq:complexshear}
\end{equation}
Thus the leading TT response is in quadrature with the strain and proportional to frequency: it has the rheological character of a purely rate-dependent shear response rather than a static shear modulus. Locally, the constitutive law can equivalently be generated by the quadratic work potential
\begin{equation}
\mathscr R_\phi[\sigma]
\equiv
\frac12\mathcal K\mathcal T\,
\sigma_{ab}\sigma^{ab},
\quad
\pi_{ab}^{(\phi)}
=
\frac{\partial\mathscr R_\phi}{\partial\sigma^{ab}},
\quad
\Wphi=2\mathscr R_\phi.
\label{eq:workpotential}
\end{equation}
We call $\mathscr R_\phi$ a work potential rather than an entropy-production potential. Indeed, in the conventional Eckart notation $\pi_{ab}=-2\eta\sigma_{ab}$ one has $\eta_{\rm eff}=-\mathcal K\mathcal T/2$ on the selected branch. The sign therefore characterizes an active effective gravitational medium, capable of doing work on the TT deformation, rather than an ordinary passive viscous fluid (see~\cite{Faraoni:2018qdr,Faraoni:2021lfc} for further details).

\subsection{Gauge invariance and second-order scope}

The appearance of a quadratic TT quantity immediately raises a perturbative question: is the work scalar itself well defined at second order, or is Eq.~\eqref{eq:Pmain} tied to a particular gauge choice?
In the present case the answer is especially simple.
Since
\begin{equation}
\bar\Wphi=0,
\qquad
\delta\Wphi=0
\label{eq:Wlow}
\end{equation}
on a shear-free background, the full second-order perturbation of $\Wphi$ is gauge invariant.
For a scalar $W$, a second-order gauge transformation is
\begin{equation}
\widetilde{\delta^{(2)}W}
=
\delta^{(2)}W
+\mathcal L_{\xi_2}\bar W
+\mathcal L_{\xi_1}^2\bar W
+2\mathcal L_{\xi_1}\delta W,
\label{eq:gauge}
\end{equation}
so Eq.~\eqref{eq:Wlow} implies
\begin{equation}
\widetilde{\delta^{(2)}\Wphi}
=
\delta^{(2)}\Wphi.
\label{eq:WGI}
\end{equation}
This is the standard second-order gauge-invariance criterion for a quantity whose background and complete first-order perturbations vanish
\cite{Bruni:1996im}.
Moreover, $\chi_{ij}^{\TT}$ is first-order gauge invariant on an FLRW background
\cite{Bardeen:1980kt,Malik:2008im}, so the TT$\times$TT coefficient in Eq.~\eqref{eq:Pmain} is separately fixed:
\begin{equation}
\frac12
\left[\delta^{(2)}\Wphi\right]_{\TT}
=
(\Wphi)_{\TT}^{[1,1]}
=
\PTT.
\label{eq:Pnorm}
\end{equation}

This does not mean that $\PTT$ is the complete second-order mechanical power. In particular, although a pure TT perturbation is trace free at first order and therefore satisfies $\delta\Theta|_{\TT}=0$, it produces a nonvanishing quadratic correction to the local volume expansion.

For the scalar-frame congruence, the expansion scalar can be written as
\begin{equation}
\Theta = \frac12 h^{ab}\mathcal L_u h_{ab} = \mathcal L_u\ln\sqrt{h},
\end{equation}
where $h$ is the determinant of the induced spatial metric. For the additive metric parametrization in Eq.~\eqref{eq:metric}, the part of the spatial determinant built purely from two first-order TT
perturbations is
\begin{equation}
\sqrt{h}
=
a^3
\left[
1-\frac{\varepsilon^2}{4}
\chi_{ij}^{\TT}\chi_{\TT}^{ij}
+O(\varepsilon^3)
\right].
\end{equation}
It follows that the corresponding TT$\times$TT contribution to the
expansion scalar is
\begin{equation}
\Theta_{\TT}^{[1,1]}
=
-\frac12
\chi_{ij}^{\TT}\dot\chi_{\TT}^{ij}.
\label{eq:Thetaquad}
\end{equation}
Thus a deformation that is trace free at linear order can nevertheless
generate an isotropic volume response at quadratic order. Equation~\eqref{eq:Thetaquad} contains only the $[1,1]$ contribution; genuine second-order lapse, shift, scalar, and spatial-metric perturbations supply additional $[2]$ terms and are not assumed to vanish.

Direct projection of the scalar effective stress shows that this quadratic expansion also induces an isotropic pressure response,
\begin{align}
(p_\phi)_{\TT}^{[1,1]}
&= \frac{\dot\phib}{12\pi\phib}\, \Theta_{\TT}^{[1,1]}, \nonumber\\
(p_\phi\Theta)_{\TT}^{[1,1]} &= \left( \bar p_\phi + \frac{\bar\Theta\dot\phib}{12\pi\phib} \right) \Theta_{\TT}^{[1,1]}.
\label{eq:pTheta}
\end{align}
This contribution is generated by the TT perturbation but is isotropic: it belongs to the pressure--expansion channel $p_\phi\Theta$. It is therefore distinct from, and does not modify, the trace-free shear-work scalar~\eqref{eq:W}.

\subsection{Exact scalar-sector balance}

Having identified the work mechanically and established its perturbative status, we will now check where it enters the scalar dynamics. This provides an independent first-law test of the interpretation and extends the linear thermodynamic-channel analysis of Ref.~\cite{Pereira:2026xog} to the leading tensor-induced nonlinear power.
In vacuum,
$\nabla_aT_{(\phi)}^{ab}=0$, and projection along $u_b$ yields
\begin{equation}
\mathscr D_u\rho_\phi+(\rho_\phi+p_\phi)\Theta+D_aq_\phi^a+2a_aq_\phi^a+\Wphi=0.
\label{eq:energyexact}
\end{equation}

Because $u^a$ is hypersurface orthogonal, we may introduce orthogonal
scalar-frame slices $t=\mathrm{const}$ such that
\begin{equation}
u_a=-N\nabla_a t,
\qquad
a_a=D_a\ln N,
\label{eq:lapse_acceleration}
\end{equation}
where $N$ is the lapse. Choosing spatial coordinates comoving with the
congruence, so that the shift vanishes and $(\partial_t)^a=Nu^a$, the
proper volume element on the orthogonal slices obeys
\begin{equation}
\partial_t(\dd V)=N\Theta\,\dd V.
\end{equation}
Equation~\eqref{eq:energyexact} therefore gives
\begin{align}
\frac{\dd}{\dd t}
\int_{\mathcal V_t}\!\rho_\phi\,\dd V
={}&
-\int_{\mathcal V_t}\!
N\left(p_\phi\Theta+\Wphi\right)\dd V
\nonumber\\
&-
\oint_{\partial\mathcal V_t}\!
Nq_\phi^a n_a\,\dd A
-\int_{\mathcal V_t}\!
N a_aq_\phi^a\,\dd V ,
\label{eq:cellfirstlaw}
\end{align}
where we used
\begin{equation}
D_a(Nq_\phi^a)
=
N D_aq_\phi^a
+
N a_aq_\phi^a.
\end{equation}
This is the exact finite-volume balance on the scalar-frame orthogonal
slices. The lapse factors are essential for a generic accelerated
congruence: when $a_a\neq0$, the lapse is spatially varying and a single
proper-time increment cannot in general be identified with the same
coordinate-time increment throughout a finite cell.

For the homogeneous FLRW background used for the pure-TT calculation, the
lapse is spatially constant and can be chosen as $N=1$ in cosmic time.
Moreover,
$\delta a_a|_{\TT}=\delta q_a^{(\phi)}|_{\TT}=0$
at first order. Hence the leading TT$\times$TT deviatoric contribution to
the scalar-cell energy balance is
\begin{equation}
\left[\frac{\dd}{\dd t}\int_{\mathcal V_t}\!\rho_\phi\,\dd V\right]_{{\rm sh},\TT}^{[1,1]}=-\int_{\mathcal V_t}\!\PTT\,\dd V,
\label{eq:cellfirstlawTT}
\end{equation}
with the isotropic pressure--expansion contribution and genuine second-order fields kept separate.

Equation~\eqref{eq:cellfirstlawTT} also makes the sign convention of the traction construction explicit. On the selected branch, $\PTT>0$ denotes positive work performed by the effective scalar stress on the TT deformation and therefore appears with a minus sign in the scalar-sector internal-energy change. The gravitational wave supplies the TT deformation that activates this constitutive work channel. In this sense the tensor mode participates in a local thermodynamic exchange with the scalar medium, without requiring a unique decomposition into autonomous scalar and gravitational-wave energies.

Furthermore, by defining
\begin{equation}
F_\phi
\equiv
\mathscr D_u\rho_\phi
+
(\rho_\phi+p_\phi)\Theta
+
D_aq_\phi^a
+
2a_aq_\phi^a,
\label{eq:F}
\end{equation}
one has then
\begin{equation}
F_\phi+\Wphi=0.
\label{eq:balance}
\end{equation}
Thus $\Wphi$ is not merely a constitutive quadratic: it is the deviatoric
stress-power term in the exact local scalar-sector energy balance. With the
traction convention adopted above, positive $\Wphi$ denotes work performed
by the scalar stress on the deformation and consequently contributes with
the opposite sign to the scalar-sector internal-energy change. The corresponding balance in the presence of minimally coupled Jordan-frame matter is given in Appendix~\ref{app:extensions}.

At second order,
\begin{equation}
\frac12
\left[\delta^{(2)}F_\phi\right]_{\TT}
=
(F_\phi)_{\TT}^{[1,1]}
+
(F_\phi)_{\TT}^{[2]}
=
-\PTT.
\label{eq:secondbalance}
\end{equation}
The right-hand side is determined by first-order TT fields, while the left-hand side includes both TT bilinears and genuine tensor-induced second-order fields.
Consequently,
$(F_\phi)_{\TT}^{[1,1]}=-\PTT$
does not follow in isolation.

Under additional Eckart--Gibbs assumptions, the same work scalar appears in a conditional entropy identity,
\begin{equation}
\calT\nabla_as_\phi^a
=
-q_a^{(\phi)}\calA^a-\Wphi,
\qquad
\calA_a=D_a\ln\calT+a_a.
\label{eq:entropy}
\end{equation}
The assumptions and derivation are summarized in Appendix~\ref{app:entropy}.

For the pure TT channel one can isolate the same conditional quadratic entropy diagnostic already anticipated in Ref.~\cite{Pereira:2026xog}. Since the first-order TT perturbation does not excite the heat flux, the TT$\times$TT shear contribution to Eq.~\eqref{eq:entropy} is
\begin{equation}
\left[\calT\nabla_a s_\phi^a\right]_{{\rm sh},\TT}^{[1,1]}=-\PTT.
\label{eq:entropyTT}
\end{equation}
Thus the gravitational theory fixes the temperature-weighted entropy-balance contribution of the shear channel even though it does not fix $\mathcal K$ and $\mathcal T$ separately. On the future-directed branch, Eq.~\eqref{eq:entropyTT} is negative because the effective shear viscosity is negative, in agreement with Ref.~\cite{Pereira:2026xog}. This is not ordinary positive viscous entropy production and should not be interpreted as a standalone second-law statement for an isolated material fluid. Genuine second-order fields and other thermodynamic channels can contribute to the complete second-order entropy balance and are not included in Eq.~\eqref{eq:entropyTT}.

Accordingly, the terminology used does not identify thermodynamic work with irreversible entropy generation. The mechanically robust statements are the traction power~\eqref{eq:W}, its cyclic work~\eqref{eq:cyclicwork}, and its appearance in the exact scalar energy balance~\eqref{eq:balance}. Equation~\eqref{eq:entropyTT} is an additional conditional statement: under the Eckart--Gibbs completion, the same shear-work channel also enters the entropy balance, without requiring an autonomous entropy for the gravitational wave itself. A microscopic or closed-system second law would require an entropy accounting for the complementary gravitational sector, which is not supplied by the present effective theory. A causal effective-thermodynamic framework~\cite{Banerjee:2026ulx} may provide some of the additional structure needed for such a more complete description.

\section{Thermodynamic exchange encoded in tensor propagation}
\label{sec:prop}

We have answered the first part of the question by identifying the local work performed on the scalar medium.
We will now address the second: is this thermodynamic exchange encoded in the propagation of the gravitational wave itself?
The key observation is that the same TT anisotropic stress identified in Ref.~\cite{Pereira:2026xog} as the non-GR tensor channel is precisely the stress entering the local mechanical power derived above.

\subsection{Direct source-power identity}

The first-order effective anisotropic stress is given by~\cite{Pereira:2026xog}
\begin{equation}
\delta\pi^{(\phi)i}{}_{j}\big|_{\TT}
=
\frac{\overline{\mathcal K\mathcal T}}{2}
\dot\chi^i{}_{j}
=
-\frac{\dot\phib}{16\pi\phib}
\dot\chi^i{}_{j}.
\label{eq:piTT}
\end{equation}
Equation~\eqref{eq:tensor_eq} can therefore be written as
\begin{equation}
\ddot\chi^i{}_{j}
+3H\dot\chi^i{}_{j}
-a^{-2}\partial^2\chi^i{}_{j}
=
16\pi\,
\delta\pi^{(\phi)i}{}_{j}\big|_{\TT}.
\label{eq:sourced}
\end{equation}
This is an algebraic rearrangement of the same scalar--tensor field equation, not the introduction of an independent matter source.
Its physical significance comes from the fact that the effective source has already been identified, independently through traction mechanics, as the stress conjugate to the scalar-frame shear.

Multiplying by $\dot\chi^j{}_{i}/(32\pi)$ gives
\begin{equation}
\frac12
\delta\pi^{(\phi)i}{}_{j}\big|_{\TT}
\dot\chi^j{}_{i}
=
\delta\pi_{ab}^{(\phi)}\big|_{\TT}
\delta\sigma^{ab}\big|_{\TT}
=
\PTT.
\label{eq:sourcepower}
\end{equation}

The effective TT source in the propagation equation performs exactly the same local shear work selected by the virtual-power construction.
The argument is now closed from both sides: traction mechanics fixes the stress--deformation power, while the field equation shows that the same stress is responsible for the modified propagation. Using Eq.~\eqref{eq:alphaM}, the work coefficient can be written
\begin{equation}
\PTT
=
-\frac{H\alpha_M}{32\pi}
\dot\chi_{ij}^{\TT}\dot\chi_{\TT}^{ij}.
\label{eq:Palpha}
\end{equation}
The same $\alpha_M$ that changes the tensor friction coefficient
$3H\rightarrow(3+\alpha_M)H$
therefore controls the local TT shear-work rate.

\subsection{GR-normalized tensor balance}

To see how this local source power appears in familiar wave-energy bookkeeping, we will now introduce the GR-normalized quadratic tensor density and flux.
Define
\begin{align}
\calE_T^{\GR}
&=
\frac{1}{64\pi}
\left[
\dot\chi_{ij}^{\TT}\dot\chi_{\TT}^{ij}
+
a^{-2}
\partial_k\chi_{ij}^{\TT}
\partial^k\chi_{\TT}^{ij}
\right],
\nonumber\\
\calF_T^i
&=
-\frac{1}{32\pi a^2}
\dot\chi_{jk}^{\TT}\partial^i\chi_{\TT}^{jk}.
\label{eq:EGR}
\end{align}
Multiplying Eq.~\eqref{eq:tensor_eq} by
$\dot\chi_{\TT}^{ij}/(32\pi)$ gives
\begin{align}
\dot{\calE}_T^{\GR}
+\partial_i\calF_T^i
={}&
-\frac{3H}{32\pi}
\dot\chi_{ij}^{\TT}\dot\chi_{\TT}^{ij}
\nonumber\\
&-
\frac{H}{32\pi a^2}
\partial_k\chi_{ij}^{\TT}
\partial^k\chi_{\TT}^{ij}
+
\PTT.
\label{eq:Ebalance}
\end{align}
The first two terms are standard FLRW redshifting.
The last term is the non-GR Planck-mass-running contribution in this normalization.
On the decreasing-$\phi$ branch it is positive: the scalar-frame sector performs positive work on the TT deformation, while the GR-normalized wave balance receives the same quantity with the opposite bookkeeping role relative to the scalar energy equation~\eqref{eq:balance}.

This does not define a unique local exchange between autonomous scalar and tensor energies.
A time-dependent normalization of the tensor action redistributes the explicit $\dot\phib$ contribution, as shown in Appendix~\ref{app:norm}.
The normalization-independent statements relevant here are the scalar-sector stress power~\eqref{eq:W}, the exact scalar balance~\eqref{eq:balance}, and the direct source-power identity~\eqref{eq:sourcepower}.

\subsection{Propagation consequences: geometric optics and standard sirens}

The source-power identity already answers the propagation part of the thermodynamic question locally.
We will now extract two further consequences of the same coefficient: a geometric-optics work law and the standard running-Planck-mass modification of gravitational-wave distances.
The previous identities are exact at the perturbative order considered, but their physical content becomes particularly transparent in geometric optics.
We will now assume that the wave period and wavelength are short compared with the background-evolution scale and the scale on which the ray bundle changes appreciably.
We denote by $\langle\cdots\rangle_{\rm cyc}$ a local phase average centered on a spacetime event, over a window large compared with the wave period but small compared with the background-evolution scale.

For a locally luminal TT wave, the leading WKB terms obey
\begin{equation}
\left\langle
a^{-2}
\partial_k\chi_{ij}^{\TT}
\partial^k\chi_{\TT}^{ij}
\right\rangle_{\rm cyc}
=
\left\langle
\dot\chi_{ij}^{\TT}\dot\chi_{\TT}^{ij}
\right\rangle_{\rm cyc}.
\label{eq:equip}
\end{equation}
Equation~\eqref{eq:EGR} then gives
\begin{equation}
\left\langle\calE_T^{\GR}\right\rangle_{\rm cyc}
=
\frac{1}{32\pi}
\left\langle
\dot\chi_{ij}^{\TT}\dot\chi_{\TT}^{ij}
\right\rangle_{\rm cyc}.
\label{eq:EWKB}
\end{equation}
Combining this relation with Eq.~\eqref{eq:Pphi}, one obtains the local work law
\begin{equation}
\frac{
\left\langle\PTT\right\rangle_{\rm cyc}
}{
\left\langle\calE_T^{\GR}\right\rangle_{\rm cyc}
}
=
-\frac{\dot\phib}{\phib}
=
-H\alpha_M.
\label{eq:Pratio}
\end{equation}
The inverse local shear-work timescale is therefore
\begin{equation}
\tau_{\rm sh}^{-1}
\equiv
\left|
\frac{\langle\PTT\rangle_{\rm cyc}}
{\langle\calE_T^{\GR}\rangle_{\rm cyc}}
\right|
=
\left|\frac{\dot\phib}{\phib}\right|
=
|H\alpha_M|.
\label{eq:timescale}
\end{equation}
For an expanding background, $H>0$, the quantity $-\alpha_M$ measures the fractional anomalous TT work per Hubble time on the selected decreasing-$\phi$ branch. The local GR-normalized relation~\eqref{eq:Pratio}, however, requires neither $H>0$ nor any additional spatial averaging within the homogeneous FLRW setting considered here. Its first equality remains meaningful at $H=0$, whereas the representation $-H\alpha_M$ assumes $H\neq0$, as specified in~\eqref{eq:alphaM}; an explicit flat-background example with $H=0$ and nonvanishing shear work is given in Appendix~\ref{app:extensions}.

The local work rate also defines a finite dimensionless accumulation along the scalar-frame evolution. For two times $t_1<t_2$ in the WKB regime, define the cumulative shear-work depth
\begin{equation}
\Xi_{\rm sh}(t_1,t_2)
\equiv
\int_{t_1}^{t_2}\!\dd t\,
\frac{\langle\PTT\rangle_{\rm cyc}}
{\langle\calE_T^{\GR}\rangle_{\rm cyc}}.
\label{eq:Xish}
\end{equation}
Using Eq.~\eqref{eq:Pratio},
\begin{equation}
\Xi_{\rm sh}(t_1,t_2)
=
-\ln\!\frac{\phib(t_2)}{\phib(t_1)}
=
-\ln\!\frac{M_*^2(t_2)}{M_*^2(t_1)}.
\label{eq:Xishintegrated}
\end{equation}
This quantity is the integrated fractional work rate associated specifically with the scalar--tensor shear channel. It is local-to-integrated thermodynamic information and does not rely on introducing a luminosity distance or on comparing gravitational and electromagnetic rays.

Cycle averaging the exact balance~\eqref{eq:Ebalance} instead gives the local transport equation
\begin{align}
\partial_t
\left\langle\calE_T^{\GR}\right\rangle_{\rm cyc}
+
\partial_i
\left\langle\calF_T^i\right\rangle_{\rm cyc}
\nonumber\\
+
4H
\left\langle\calE_T^{\GR}\right\rangle_{\rm cyc}
={}&
-\frac{\dot\phib}{\phib}
\left\langle\calE_T^{\GR}\right\rangle_{\rm cyc}.
\label{eq:Eredshift_general}
\end{align}
The flux divergence in Eq.~\eqref{eq:Eredshift_general} is part of the leading transport problem and should not be removed by cycle averaging alone.
For a localized packet or a spherical wave it contains the geometric spreading of the radiation.

There are two common situations in which one can remove this term, but they should be distinguished.
For a homogeneous plane wave one has locally
\begin{equation}
\partial_i
\left\langle\calF_T^i\right\rangle_{\rm cyc}=0.
\label{eq:plane_wave_flux}
\end{equation}
Alternatively, after performing the same local cycle average used in Eq.~\eqref{eq:equip}, let
$\langle\cdots\rangle_{{\rm cyc},V}$ denote an additional spatial average over a comoving volume $V$.
Then the divergence term obeys
\begin{equation}
\left\langle
\partial_i\calF_T^i
\right\rangle_{{\rm cyc},V}
=
\frac{1}{V}
\oint_{\partial V}
\left\langle\calF_T^i\right\rangle_{\rm cyc}
\,\dd\Sigma_i,
\label{eq:volume_flux}
\end{equation}
which vanishes for periodic boundaries or vanishing net boundary flux.
The WKB equipartition relation~\eqref{eq:equip} is understood to be applied before this spatial average.
In either of these specializations the averaged energy obeys
\begin{equation}
\frac{\dd}{\dd t}
\left\langle\calE_T^{\GR}\right\rangle
+
4H
\left\langle\calE_T^{\GR}\right\rangle
=
-\frac{\dot\phib}{\phib}
\left\langle\calE_T^{\GR}\right\rangle,
\label{eq:Eredshift}
\end{equation}
where the brackets now denote either the local cycle average for a homogeneous plane wave or the combined cycle-and-volume average $\langle\cdots\rangle_{{\rm cyc},V}$.
It follows that
\begin{equation}
\left\langle\calE_T^{\GR}\right\rangle
\propto
a^{-4}\phib^{-1}.
\label{eq:Escale}
\end{equation}

Furthermore, there is also a useful complementary invariant in the geometric-optics regime. With the action normalization introduced in Appendix~\ref{app:norm},
$\calE_T^{\rm act}=\phib\calE_T^{\GR}$, Eq.~\eqref{eq:Escale} implies
$\langle\calE_T^{\rm act}\rangle\propto a^{-4}$. Since $\omega\propto a^{-1}$, the comoving wave action
\begin{equation}
\mathscr J_T
\equiv
\frac{a^3\langle\calE_T^{\rm act}\rangle}{\omega}
=
\frac{a^3\phib\langle\calE_T^{\GR}\rangle}{\omega}
\label{eq:waveaction}
\end{equation}
is conserved at leading WKB order in the plane-wave/vanishing-net-flux settings used in deriving Eq.~\eqref{eq:Escale}. Thus the shear-work term should not be conflated with stochastic absorption or particle production: at geometric-optics order it is an adiabatic parametric work exchange associated with the running tensor normalization, while the action-normalized wave action remains conserved. This provides a thermodynamic interpretation of the process that is independent of entropy production.

Since the physical frequency redshifts as $a^{-1}$, the background-dependent part of the plane-wave or Fourier-mode amplitude is
\begin{equation}
A_\lambda
\propto
\frac{1}{a\sqrt{\phib}}.
\label{eq:amp}
\end{equation}

The last relation should not be confused with the complete amplitude of radiation from a localized source.
For a ray bundle, Eq.~\eqref{eq:Eredshift_general} also transports the usual geometric dilution through
$\partial_i\langle\calF_T^i\rangle_{\rm cyc}$.
For luminal gravitational and electromagnetic rays on the same background, this geometrical factor is common to both sectors, while the additional gravitational-wave normalization is controlled by the running Planck mass.
Consequently,
\begin{equation}
\frac{d_L^{\rm GW}(z)}{d_L^{\rm EM}(z)}
=
\frac{M_*(0)}{M_*(z)}
=
\sqrt{\frac{\bar\phi_0}{\bar\phi(z)}}.
\label{eq:dL}
\end{equation}
Equivalently,
\begin{equation}
\ln
\frac{d_L^{\rm GW}}{d_L^{\rm EM}}
=
\frac12
\int_0^z
\frac{\alpha_M(z')}{1+z'}\,\dd z'.
\label{eq:dLint}
\end{equation}
This is the standard running-Planck-mass propagation law
\cite{Belgacem:2017ihm,Belgacem:2018,Lagos:2019kds,Amendola:2017ovw}.
The new statement for Eqs.~\eqref{eq:dL}--\eqref{eq:dLint} is instead that the same coefficient entering that known propagation law has already been identified independently as the local scalar-frame traction-work coefficient, and that its accumulated local effect can be written as the shear-work depth~\eqref{eq:Xishintegrated}. In this sense, the standard-siren relation is an integrated observational representation of the local work channel, rather than the origin of the thermodynamic interpretation.

\section{How general is the thermodynamic shear-excitation mechanism?}
\label{sec:families}

Having answered the thermodynamic-exchange question in Bergmann--Wagoner gravity, we will now ask how much of the mechanism is specific to that theory. The relevant issue is not merely whether another theory modifies tensor damping, but whether it also supplies an independently defined internal stress whose power can be identified with the tensor deformation. This question is nontrivial. The recent analysis of $F(\Phi,X)R$ theories shows that generalized nonminimal couplings can generate heat-flux components that are not captured by the simplest Eckart acceleration law \cite{Pereira:2025dmk}, while tensor--multiscalar gravity can contain additional field-space thermal directions even in the coupling-comoving frame \cite{Pereira:2026hpi}. Accordingly, we will distinguish carefully between the {shear-work} coefficient relevant for TT propagation and the more general heat-sector variables. In every extension we will remain in the positive tensor-kinetic sector: $F>0$ for a nonminimal coupling, $f_R>0$ in metric $f(R)$ gravity, $G_4>0$ in the luminal Horndeski subclass, and $M_T^2>0$ in the generic tensor action. The Bergmann--Wagoner result will first be extended to closely related one-field families and then to viable Horndeski gravity.

\subsection{Brans--Dicke, nonminimal coupling, and metric \texorpdfstring{$f(R)$}{f(R)} gravity}

For Brans--Dicke gravity,
$\omega(\phi)=\omega_{\rm BD}$ and the same equations apply directly.
The explicit TT work coefficient is independent of $\omega_{\rm BD}$ and $V$:
\begin{equation}
\mathcal P_{\TT,{\rm BD}}^{[1,1]}
=
-\frac{\dot\phib}{32\pi\phib}
\dot\chi_{ij}^{\TT}\dot\chi_{\TT}^{ij}.
\label{eq:PBD}
\end{equation}
The parameter $\omega_{\rm BD}$ and the potential affect the work indirectly through the background solution $\phib(t)$.

For a nonminimally coupled scalar with gravitational coupling $F(\varphi)>0$,
\begin{equation}
S\supset
\frac{1}{16\pi}
\int\!\dd^4x\sqrt{-g}\,
F(\varphi)R,
\end{equation}
the replacement $\phi\rightarrow F(\varphi)$ gives
\begin{equation}
\mathcal P_{\TT,F}^{[1,1]}
=
-\frac{\dot F}{32\pi F}
\dot\chi_{ij}^{\TT}\dot\chi_{\TT}^{ij}.
\label{eq:PF}
\end{equation}
Therefore the work channel is controlled by the logarithmic running of the nonminimal coupling rather than by the choice of scalar field coordinate.

On a regular Legendre-transform branch, metric $f(R)$ gravity is dynamically equivalent to the $\omega=0$ scalar--tensor sector with
$\phi=f_R\equiv\dd f/\dd R$
\cite{Sotiriou:2008rp,Faraoni:2018qdr}.
We restrict to $f_R>0$, which is the positive tensor-kinetic sector corresponding to $\phi>0$.
Consequently,
\begin{equation}
\mathcal P_{\TT,f(R)}^{[1,1]}
=
-\frac{\dot f_R}{32\pi f_R}
\dot\chi_{ij}^{\TT}\dot\chi_{\TT}^{ij},
\label{eq:PfR}
\end{equation}
and
\begin{equation}
\alpha_M^{f(R)}
=
\frac{\dot f_R}{Hf_R}.
\label{eq:alphaMfR}
\end{equation}
This provides an immediate thermodynamic interpretation of the running tensor normalization in metric $f(R)$ cosmology.

\subsection{Viable luminal Horndeski gravity}

A more significant extension occurs in the viable luminal Horndeski subclass
\cite{Horndeski:1974wa,Giusti:2021sku,Miranda:2024dhw},
\begin{align}
\mathcal L_{\rm H}
={}&
G_2(\varphi,X_{\rm H})
-
G_3(\varphi,X_{\rm H})\Box\varphi
+
G_4(\varphi)R,
\label{eq:HorndeskiL}
\\
X_{\rm H}
\equiv{}&
-\frac12\nabla_a\varphi\nabla^a\varphi.
\label{eq:HorndeskiX}
\end{align}
This is precisely the Horndeski structure that remains naturally luminal without cancellations involving quartic or quintic derivative couplings after the GW170817 constraint
\cite{Baker:2017hug,Creminelli:2017sry,Ezquiaga:2017ekz,Sakstein:2017xjx}.
We will also assume $G_4>0$, so that $M_*^2=2G_4>0$ and the tensor kinetic term has the correct sign.

In the effective Einstein description used in the Horndeski thermodynamic literature, the anisotropic stress satisfies
\cite{Giusti:2021sku,Nucamendi:2019uen}
\begin{equation}
\pi_{ab}^{\rm eff}
=
\frac{G_{4\varphi}\sqrt{2X_{\rm H}}}{G_4}\,
\sigma_{ab}.
\label{eq:HorndeskiPiRaw}
\end{equation}
To compare with the normalization $G_{ab}=8\pi T_{ab}$ used in the present paper, define
$\widehat T_{ab}^{\rm eff}\equiv T_{ab}^{\rm eff}/(8\pi)$.
Then
\begin{equation}
\widehat\pi_{ab}^{\rm eff}
=
\frac{G_{4\varphi}\sqrt{2X_{\rm H}}}{8\pi G_4}\,
\sigma_{ab}.
\label{eq:HorndeskiPi}
\end{equation}

For a homogeneous future-directed scalar gradient,
$\sqrt{2X_{\rm H}}=-\dot\varphi$.
Since
\begin{equation}
M_*^2=2G_4,
\qquad
\alpha_M
=
\frac{\dot G_4}{HG_4},
\label{eq:Halpha}
\end{equation}
the shear coefficient becomes
\begin{equation}
\frac{G_{4\varphi}\sqrt{2X_{\rm H}}}{8\pi G_4}
=
-\frac{\dot G_4}{8\pi G_4}
=
-\frac{H\alpha_M}{8\pi}.
\label{eq:Hcoeff}
\end{equation}
The tensor-induced Horndeski shear work is therefore
\begin{equation}
\mathcal P_{\TT,{\rm H}}^{[1,1]}
=
-\frac{H\alpha_M}{32\pi}
\dot\chi_{ij}^{\TT}\dot\chi_{\TT}^{ij}.
\label{eq:PH}
\end{equation}
This is the same structural relation obtained in Bergmann--Wagoner gravity.
In other words, the work/damping correspondence survives even though the full effective thermodynamic sector of Horndeski gravity is richer.

The Horndeski extension also reveals an important distinction that is hidden in the simpler theory.
The effective heat channel contains the combination
\cite{Giusti:2021sku}
\begin{equation}
(\mathcal K\mathcal T)_{\rm H}
\propto
\frac{\sqrt{2X_{\rm H}}}{G_4}
\left(
G_{4\varphi}-X_{\rm H}G_{3X}
\right),
\label{eq:HKT}
\end{equation}
whereas the anisotropic stress in Eq.~\eqref{eq:HorndeskiPiRaw} depends only on $G_{4\varphi}$.
Thus the braiding function $G_3$ contributes explicitly to the effective heat sector but not to the TT shear stress.
The tensor work channel therefore isolates the nonminimal-coupling/running-Planck-mass sector:
$G_3$ enters the explicit heat-transport coefficient, whereas $G_4(\varphi)$ controls the explicit TT shear-work/damping coefficient.
This statement concerns the constitutive coefficients themselves; $G_3$ may still influence $\dot G_4$ indirectly through the background dynamics.
Accordingly, beyond Bergmann--Wagoner gravity one should not identify the heat-sector product $\mathcal K\mathcal T$ with the shear-response coefficient without qualification.
This separation is conceptually consistent with the more general lesson of Ref.~\cite{Pereira:2025dmk}: once the nonminimal coupling depends on additional kinetic structure, the heat channel may acquire transverse pieces that are independent of the simple inertial acceleration term.

This separation is consistent with the broader effective-field-theory description of Horndeski perturbations, in which $\alpha_M$ tracks the running Planck mass while braiding is encoded independently by $\alpha_B$
\cite{Bellini:2014fua}.
It also extends the fluid result of Ref.~\cite{Miranda:2022wkz}: in the subclass admitting a simple Newtonian/Eckart-like anisotropic stress, the same linear shear coefficient that gives the fluid work also appears in luminal tensor damping.

\subsection{A criterion for more general Horndeski and DHOST tensor sectors}

The previous Horndeski result relies on an independently established constitutive stress law.
This qualification becomes even more important in broader scalar--tensor families, where the effective medium may contain additional spatial or field-space directions
\cite{Pereira:2026hpi}.
We will therefore formulate the extension as a criterion rather than assume that anomalous damping is automatically thermodynamic.
For a more general single-tensor-sector theory, consider the quadratic action
\begin{align}
S_T^{(2)}
=
\frac18
\int\!\dd^4x\,a^3M_T^2(t)
\Big[
&\dot\chi_{ij}^{\TT}\dot\chi_{\TT}^{ij}
\nonumber\\
&-
c_T^2(t)a^{-2}
\partial_k\chi_{ij}^{\TT}
\partial^k\chi_{\TT}^{ij}
\Big].
\label{eq:genericTaction}
\end{align}
We restrict to $M_T^2(t)>0$ so that the tensor is not a ghost.
When we refer to a stable propagating tensor sector we additionally assume
$c_T^2(t)>0$, excluding a short-wavelength tensor gradient instability.
No scalar-sector stability conditions are needed for the stress-power identity derived here.

Its equation of motion can be written with mixed spatial components as
\begin{equation}
\ddot\chi^i{}_{j}
+
\left(
3H+\nu_T
\right)\dot\chi^i{}_{j}
-
c_T^2a^{-2}\partial^2\chi^i{}_{j}
=
0.
\label{eq:genericTeq}
\end{equation}
Here the anomalous tensor-friction rate is
\begin{align}
\nu_T
&\equiv
\frac{\partial_t(M_T^2)}{M_T^2}
=
\frac{\dd}{\dd t}\ln M_T^2
=
2\frac{\dot M_T}{M_T}.\label{eq:nuTdef}
\end{align}
Relative to the GR propagation operator, one then has
\begin{align}
\ddot\chi^i{}_{j}
+3H\dot\chi^i{}_{j}
-a^{-2}\partial^2\chi^i{}_{j}
={}&
-\nu_T\dot\chi^i{}_{j}
\nonumber\\
&+
(c_T^2-1)a^{-2}\partial^2\chi^i{}_{j}.
\label{eq:genericSource}
\end{align}
Keeping the same mixed-index normalization used in Sec.~\ref{sec:prop}, we may define the corresponding normalized effective TT stress by
\begin{align}
16\pi\,\delta\pi^i{}_{j,\rm eff}
\equiv{}
-\nu_T\dot\chi^i{}_{j}+
(c_T^2-1)a^{-2}\partial^2\chi^i{}_{j}.
\label{eq:genericEffectiveStress}
\end{align}
Equivalently,
\begin{align}
\delta\pi^i{}_{j,\rm eff}=\frac{1}{16\pi}\Big[
-\nu_T\dot\chi^i{}_{j}
+
(c_T^2-1)a^{-2}\partial^2\chi^i{}_{j}
\Big].
\label{eq:genericEffectiveStressExplicit}
\end{align}
For comparison, covariant spatial components obey
$\delta\pi_{ij}^{\rm eff}=a^2\delta_{ik}\,\delta\pi^k{}_{j,\rm eff}$ at this perturbative order, with the index converted using the background spatial metric.

Multiplying Eq.~\eqref{eq:genericSource} by
$\dot\chi^j{}_{i}/(32\pi)$ gives the propagation-side source power
\begin{align}
\mathcal P_{\rm prop}
={}&
-\frac{\nu_T}{32\pi}
\dot\chi_{ij}^{\TT}\dot\chi_{\TT}^{ij}
\nonumber\\
&+
\frac{c_T^2-1}{32\pi a^2}
\dot\chi_{ij}^{\TT}\partial^2\chi_{\TT}^{ij}.
\label{eq:genericPower}
\end{align}
The first term is local and quadratic in the deformation rate; the second involves spatial gradients and is not of the simple shear-square form. 

This gives a useful criterion.
For a luminal tensor sector, $c_T=1$, an anomalous damping term generated solely by $M_T^2(t)$ has the propagation-side power
\begin{equation}
\mathcal P_{\rm prop}^{(c_T=1)}
=
-\frac{\partial_t(M_T^2)}{32\pi M_T^2}
\dot\chi_{ij}^{\TT}\dot\chi_{\TT}^{ij}.
\label{eq:genericLuminal}
\end{equation}
It acquires the same {thermodynamic} interpretation as the present scalar--tensor result if, independently, the effective gravitational medium admits a local constitutive anisotropic stress
\begin{align}
\delta\pi_{ab}^{\rm eff}\big|_{\TT}
&=
C_\sigma(t)\,
\delta\sigma_{ab}\big|_{\TT},
\label{eq:criterion_stress}
\\
C_\sigma
&=
-\frac{\partial_t(M_T^2)}
{8\pi M_T^2}.
\label{eq:criterion}
\end{align}
When this condition holds,
$\pi_{ab}^{\rm eff}\sigma^{ab}$
reproduces Eq.~\eqref{eq:genericLuminal}.
When it does not hold, anomalous damping alone is insufficient to claim a thermodynamic shear-work law. 

A concrete counterexample shows why the condition is nontrivial. Consider the quartic Horndeski model with $G_5=0$ and
\begin{equation}
G_4(\varphi,X_{\rm H})=F(\varphi)+\beta X_{\rm H},
\qquad \beta\neq0,
\label{eq:quarticexample}
\end{equation}
with the standard quartic Horndeski counterterm understood. The tensor kinetic and gradient coefficients are \cite{Kobayashi:2011nu}
\begin{equation}
M_T^2
=2\left(F-\beta X_{\rm H}\right),
\qquad
c_T^2
=
\frac{F+\beta X_{\rm H}}
{F-\beta X_{\rm H}}.
\label{eq:quartictensor}
\end{equation}
Whenever $F-\beta X_{\rm H}$ evolves, the tensor equation has anomalous damping,
\begin{equation}
\nu_T
=
\frac{\dd}{\dd t}
\ln\!\left[2(F-\beta X_{\rm H})\right]
\neq0,
\label{eq:quarticdamping}
\end{equation}
but the propagation-side effective stress also contains
\begin{equation}
\frac{c_T^2-1}{16\pi a^2}\partial^2\chi^i{}_j
=
\frac{\beta X_{\rm H}}
{8\pi a^2(F-\beta X_{\rm H})}
\partial^2\chi^i{}_j.
\label{eq:quarticgradient}
\end{equation}
For arbitrary TT configurations this spatial-gradient term cannot be represented by a local constitutive law $\delta\pi_{ab}=C_\sigma(t)\delta\sigma_{ab}$, even though anomalous damping is present. The model is strongly constrained as a late-time cosmology by the gravitational-wave speed bound, but it provides an explicit theoretical counterexample to the implication ``anomalous damping $\Rightarrow$ thermodynamic shear medium.'' Conversely, even in luminal theories, the existence of a running tensor normalization is not sufficient to establish a complete Eckart medium: the $F(\Phi,X)R$ example of Ref.~\cite{Pereira:2025dmk} shows that kinetic nonminimal couplings can obstruct the global heat-flux constitutive closure. These examples separate propagation kinematics from thermodynamic constitutive structure.

Hence, these results are important for beyond-Horndeski and degenerate higher-order scalar--tensor (DHOST) theories
\cite{Langlois:2015cwa,Langlois:2017dyl,Nucamendi:2019uen}.
The post-GW170817 viable sectors can be arranged so that $c_T=1$, but their effective-fluid structures can contain additional operators and need not reduce to the Newtonian/Eckart form of Eq.~\eqref{eq:criterion}.
Equation~\eqref{eq:criterion} therefore identifies what must be checked in each family: the propagation-side damping coefficient is necessary data, but a local thermodynamic interpretation requires an independently derived stress--deformation law. This criterion is the natural extension of the question posed in the Introduction: modified propagation becomes a thermodynamic shear excitation only when the theory itself supplies the internal constitutive medium on which the tensor mode performs work.

Moreover, the two terms in Eq.~\eqref{eq:genericPower} have different mechanical characters.  Writing the tensor strain as $\epsilon^i{}_j=\chi^i{}_j/2$ once more, the propagation-side effective stress becomes
\begin{equation}
\delta\pi^i{}_{j,\rm eff}
=
-\frac{\nu_T}{8\pi}\,\dot\epsilon^i{}_j
+
\frac{c_T^2-1}{8\pi a^2}\,\partial^2\epsilon^i{}_j.
\label{eq:genericRheology}
\end{equation}
For a Fourier mode with comoving wavenumber $k$, the second term is proportional to the strain itself, whereas the first is proportional to the strain rate.  In a local WKB cell in which the background coefficients are frozen over one period,
\begin{equation}
\oint\!\dd\tau\,
\frac{c_T^2-1}{8\pi a^2}
(\partial^2\epsilon^i{}_j)\dot\epsilon^j{}_i
=0,
\label{eq:reactivecycle}
\end{equation}
up to a boundary term, while
\begin{equation}
\Delta w_{\rm fr}^{\rm cyc}
=
-\frac{\nu_T}{8\pi}
\int_{\rm cyc}\!\dd\tau\,
\dot\epsilon^i{}_j\dot\epsilon^j{}_i .
\label{eq:frictioncycle}
\end{equation}
Thus a modified propagation speed is, at this level, a reactive/storage-like contribution, whereas anomalous friction is the part capable of producing nonzero closed-cycle work.  This propagation-side decomposition is only a diagnostic: it becomes a thermodynamic rheology only if the same stress follows independently from the constitutive structure of a gravitational medium.  This distinction is precisely why Eq.~\eqref{eq:criterion} is an additional physical requirement rather than a definition.

\section{Discussion and outlook}
\label{sec:discussion}

The question posed at the outset of this paper admits a precise answer in Jordan-frame scalar--tensor gravity. A propagating TT mode can participate in a local thermodynamic process without being assigned an autonomous entropy or a unique local energy of its own. The reason is that the scalar gradient supplies an internal gravitational medium with a preferred congruence, constitutive stress, and exact balance law. The tensor perturbation deforms that medium in pure shear at first order, and the corresponding anisotropic stress performs the local work identified by Eqs.~\eqref{eq:Pmech2} and~\eqref{eq:W}. Its leading TT contribution is Eq.~\eqref{eq:Pmain}. This is the sense in which the gravitational wave acts as a thermodynamic shear excitation.

The result in question also answers the second part of the question. The same effective anisotropic stress that performs the local work is the source that reproduces the non-GR tensor term in Eq.~\eqref{eq:sourced}, and the equality of the two powers is expressed by Eq.~\eqref{eq:sourcepower}. Thus the local exchange is not an interpretation imposed after inspecting the wave equation: the stress is first selected independently by traction mechanics and only then shown to be the same stress that controls tensor propagation. The scalar-sector balance~\eqref{eq:balance} and the propagation-side identity~\eqref{eq:sourcepower} are complementary descriptions of one scalar--tensor coupling.

The local work construction also leads to several results that clarify what is, and what is not, contained in the TT shear channel. The traction derivation remains valid for the accelerated scalar-gradient congruence and selects the trace-free stress power before the scalar--tensor constitutive relation is imposed. For a TT perturbation this gives the leading quadratic work~\eqref{eq:Pmain}, while Eqs.~\eqref{eq:WGI}--\eqref{eq:Pnorm} show that the corresponding second-order scalar is gauge invariant. At the same order the tensor perturbation also generates the separate isotropic response~\eqref{eq:Thetaquad}--\eqref{eq:pTheta}; hence $\PTT$ is the complete TT$\times$TT deviatoric work, but not the complete tensor-induced mechanical response. The nonvanishing closed-cycle work~\eqref{eq:cyclicwork} and the frequency-space relation~\eqref{eq:complexshear} further show that this channel is rate dependent rather than conservative-elastic. Most importantly, Eq.~\eqref{eq:cellfirstlaw} places the same quantity directly in the exact lapse-weighted finite-volume balance of the scalar medium, while Eq.~\eqref{eq:cellfirstlawTT} shows that the pure-TT shear contribution enters the scalar-sector energy change as $-\PTT$. Under the additional Eckart--Gibbs closure, Eq.~\eqref{eq:entropyTT} then identifies the corresponding conditional contribution to the entropy balance, without promoting it to ordinary positive viscous entropy production.

The propagation analysis provides an independent description of the same coupling. Equation~\eqref{eq:sourcepower} shows that the stress selected by local traction mechanics is precisely the stress responsible for the non-GR tensor term, while Eq.~\eqref{eq:Ebalance} displays its contribution in GR-normalized tensor-energy bookkeeping. Appendix~\ref{app:norm} shows at the same time why this bookkeeping should not be interpreted as a unique exchange between separately defined scalar and tensor energies. In geometric optics, Eq.~\eqref{eq:Pratio} relates the GR-normalized fractional work rate to the logarithmic running of the effective Planck mass, and Eq.~\eqref{eq:Xishintegrated} gives the corresponding accumulated diagnostic along the evolution. The transport equation~\eqref{eq:Eredshift_general} keeps this local effect distinct from geometrical dilution, while Eq.~\eqref{eq:waveaction} shows that the action-normalized comoving wave action is conserved at leading WKB order. The familiar standard-siren relations~\eqref{eq:dL}--\eqref{eq:dLint} therefore arise at the end of this chain as the integrated propagation manifestation of the same running coupling, rather than as a new propagation prediction.

The broader extensions in Sec.~\ref{sec:families} show which parts of this structure survive beyond the original Bergmann--Wagoner setting. Brans--Dicke gravity, ordinary nonminimal coupling, and metric $f(R)$ gravity retain the same logarithmic-running form of the TT work coefficient. In viable luminal Horndeski gravity, the shear channel is controlled explicitly by the running of $G_4$, whereas the heat sector also contains the independent braiding contribution $G_{3X}$; the thermodynamic heat and TT shear channels therefore need not be governed by the same coefficient. For a generic tensor sector, Eqs.~\eqref{eq:genericPower}--\eqref{eq:criterion} give a sufficient condition for recovering the same local shear-work interpretation and make clear that anomalous propagation alone does not establish an underlying thermodynamic medium. The quartic-Horndeski example~\eqref{eq:quarticexample}--\eqref{eq:quarticgradient} illustrates the failure of the broader inference from anomalous damping to a shear-only constitutive stress, while Eqs.~\eqref{eq:reactivecycle}--\eqref{eq:frictioncycle} distinguish the reactive propagation-speed contribution from the rate-dependent term capable of producing nonzero closed-cycle work. The flat-background and matter examples in Appendix~\ref{app:extensions} further show that the local shear-work law does not rely on cosmological expansion and how it is embedded in the scalar balance when ordinary Jordan-frame matter is present.

Equation~\eqref{eq:Pmain} isolates the complete TT$\times$TT deviatoric stress power, not the complete second-order scalar response; isotropic pressure--expansion work and genuine second-order fields remain present. The entropy relation~\eqref{eq:entropy} is conditional on the additional Eckart--Gibbs assumptions summarized in Appendix~\ref{app:entropy}, and the separate $\mathcal K$ and $\mathcal T$ are not fixed by the gravitational theory. We therefore distinguish throughout between the unconditional {thermodynamic work process}---a mechanically selected stress power in an independently established effective thermodynamic medium, entering its first-law balance---and the stronger notion of irreversible entropy production. The latter is not established here. The effective-fluid description is Jordan-frame specific, and Appendix~\ref{app:norm} shows that assigning an opposite term to a separately normalized local tensor energy is convention dependent. None of these qualifications affects the local virtual-power statement, the cyclic work, or the direct source--power identity.

In the Bergmann--Wagoner sector the mechanism closes smoothly in the regular GR equilibrium limit discussed in Appendix~\ref{app:extensions}: the scalar-mediated thermodynamic shear channel disappears while the spin--2 radiative degree of freedom remains. Scalar--tensor gravity therefore supplies a controlled classical prototype in which gravitational radiation participates in a genuine local thermodynamic work process through an internal gravitational constitutive sector. The additional results of this paper show how that prototype is encoded in wave propagation, how it connects to standard-siren observables, and what must survive for the interpretation to extend to broader modified-gravity theories. A separate and more fundamental question remains open: whether general relativity itself admits an analogous internal geometric, semiclassical, or quantum channel through which propagating spin--2 modes can acquire a comparable thermodynamic role.

\section*{Data Availability}
No data were created or analyzed in this study.

\begin{acknowledgments}
We thank Jess Rutschi for helpful comments. We also thank the anonymous referee for the valuable suggestions provided, which have greatly contributed to enhancing the quality of this manuscript.
This work was supported by Funda\c{c}\~ao para a Ci\^encia e a Tecnologia (FCT)
through national funds under research grant UID/04434/2025
(DOI 10.54499/UID/04434/2025).
\end{acknowledgments}

\appendix

\section{Flat background, matter, and closure of the shear-work channel}
\label{app:extensions}

For $\omega=V=0$,
\begin{equation}
g_{ab}=\eta_{ab},
\qquad
\phib=\phi_0-\mu t>0,
\qquad
\mu>0,
\end{equation}
is an exact vacuum background.
Then
\begin{equation}
\overline{\mathcal K\mathcal T}
=
\frac{\mu}{8\pi\phib},
\qquad
\PTT
=
\frac{\mu}{32\pi\phib}
\dot\chi_{ij}^{\TT}\dot\chi_{\TT}^{ij}.
\end{equation}
The linear tensor equation becomes
\begin{equation}
\ddot\chi_{ij}^{\TT}
+
\frac{\dot\phib}{\phib}
\dot\chi_{ij}^{\TT}
-
\nabla^2\chi_{ij}^{\TT}
=0.
\end{equation}
At second order the TT wave sources genuine scalar and metric corrections; the displayed work remains the isolated $[1,1]$ deviatoric term.

With minimally coupled Jordan-frame matter,
\begin{align}
G_{ab}
&=
\frac{8\pi}{\phi}T_{ab}^{(m)}
+8\pi T_{ab}^{(\phi)},
\nonumber\\
\nabla_aT_{(\phi)}^{ab}
&=
\frac{T_{(m)}^{ab}\nabla_a\phi}{\phi^2},
\end{align}
and the scalar balance becomes
\begin{equation}
F_\phi+\Wphi
=
-\frac{X}{\phi^2}\rho_m^{(u)},
\qquad
\rho_m^{(u)}=T_{ab}^{(m)}u^au^b.
\end{equation}
For homogeneous perfect matter without TT anisotropic stress, the local TT work law keeps the same form with matter-dependent background quantities.

Finally, within the Bergmann--Wagoner sector, for $\phi\rightarrow\phi_0>0$ and $\mathcal K\mathcal T\rightarrow0$,
\begin{equation}
q_a^{(\phi)},
\quad
\pi_{ab}^{(\phi)},
\quad
\PTT
\rightarrow0,
\end{equation}
while the tensor action tends to its GR form with constant normalization. This Bergmann--Wagoner GR-equilibrium limit removes the scalar-mediated shear exchange and anomalous damping, not the spin--2 radiative degree of freedom.

\section{Conditional entropy balance}
\label{app:entropy}

For completeness, we will now state the additional assumptions required to obtain the entropy identity quoted in the main text. Introduce an auxiliary number density $n_\phi$ satisfying
\begin{equation}
\nabla_a(n_\phi u^a)=0
\end{equation}
and a specific entropy $\varsigma_\phi$ obeying the Gibbs relation
\begin{equation}
\calT\,\dd\varsigma_\phi
=
\dd\left(\frac{\rho_\phi}{n_\phi}\right)
+
p_\phi\,
\dd\left(\frac1{n_\phi}\right).
\end{equation}
Along the scalar congruence,
\begin{equation}
n_\phi\calT\,\mathscr D_u\varsigma_\phi
=
\mathscr D_u\rho_\phi
+
(\rho_\phi+p_\phi)\Theta.
\end{equation}
With the Eckart current
\begin{equation}
s_\phi^a
=
n_\phi\varsigma_\phi u^a
+\frac{q_\phi^a}{\calT},
\end{equation}
and the exact energy equation~\eqref{eq:energyexact},
one obtains Eq.~\eqref{eq:entropy}.
No particle-production, reaction, or chemical-potential terms are included.
The gravitational construction fixes only the product $\mathcal K\mathcal T$.
These assumptions are why the entropy interpretation is secondary to the local mechanical work law.

\section{Normalization of the tensor-amplitude balance}
\label{app:norm}

We will now make explicit why the attribution of a separate local source to the tensor energy is normalization dependent.
Define the action-normalized quadratic quantities
\begin{equation}
\calE_T^{\rm act}
\equiv
\phib\,\calE_T^{\GR},
\qquad
\calF_{\rm act}^i
\equiv
\phib\,\calF_T^i.
\end{equation}
Using Eq.~\eqref{eq:Ebalance},
\begin{align}
\dot{\calE}_T^{\rm act}
+\partial_i\calF_{\rm act}^i
={}&
-\frac{3H\phib}{32\pi}
\dot\chi_{ij}^{\TT}\dot\chi_{\TT}^{ij}
\nonumber\\
&-
\frac{H\phib}{32\pi a^2}
\partial_k\chi_{ij}^{\TT}
\partial^k\chi_{\TT}^{ij}
\nonumber\\
&+
\frac{\dot\phib}{64\pi}
\left[
a^{-2}
\partial_k\chi_{ij}^{\TT}
\partial^k\chi_{\TT}^{ij}
-
\dot\chi_{ij}^{\TT}\dot\chi_{\TT}^{ij}
\right].
\label{eq:actionbalance}
\end{align}
Thus the isolated $+\PTT$ term of the GR-normalized balance is redistributed when the time-dependent kinetic coefficient is absorbed into the energy definition.
In geometric optics the final bracket averages to zero at leading order.
This is why the direct stress-power identity, rather than a putative local conservation law between separately normalized scalar and tensor energies, is the robust statement.

\bibliographystyle{apsrev4-2}
\bibliography{biblio}

\end{document}